\documentclass[iop]{emulateapj}
\usepackage{amsmath}
\RequirePackage{lineno}
\usepackage{graphicx}
\graphicspath{{Img/}}

\usepackage{enumerate}
\usepackage{tabularx}
\newcommand{\chisq}{$\chi^2$}

\newcommand{\ee}[1]{$\times$10$^{#1}$}
\newcommand{\cm}[1]{~cm$^{#1}$}

\newcommand{\msun}{M$_{\odot}$}

\newcommand{\ls}{{_{\sim}\atop^{<}}}
\newcommand{\Mdot}{ \dot M}

\shorttitle{The ultralong GRB130925A} \shortauthors{L. Piro et
al.}

\begin{document}
\title{A HOT COCOON IN THE ULTRALONG GRB~130925A:
\mbox{HINTS OF A POPIII-LIKE PROGENITOR IN A LOW DENSITY WIND ENVIRONMENT}}

\author{Luigi Piro\altaffilmark{1}, Eleonora Troja\altaffilmark{2}, Bruce Gendre\altaffilmark{3}, Gabriele Ghisellini\altaffilmark{4},
 Roberto Ricci\altaffilmark{5}, Keith Bannister\altaffilmark{6,7} , Fabrizio Fiore\altaffilmark{8} , Lauren~A.~Kidd\altaffilmark{2},
 Silvia Piranomonte\altaffilmark{8}, Mark H. Wieringa\altaffilmark{9}}

\altaffiltext{1}{INAF- Istituto Astrofisica e Planetologia
Spaziali, Via Fosso Cavaliere 100, I-00133, Rome, Italy}
\altaffiltext{2}{NASA/GSFC, Greenbelt, MD 20771, USA}
\altaffiltext{3}{ARTEMIS, UMR 7250, Boulevard de l' Observatoire,
Nice, Cedex 4, France} \altaffiltext{4}{ INAF-Osservatorio
Astronomico di Brera, via E. Bianchi 46, I-23807 Merate (LC),
Italy } \altaffiltext{5}{(INAF-Istituto di Radioastronomia, Via
Gobetti 101, I-40129 Bologna, Italy} \altaffiltext{6}{ CSIRO
Astronomy and Space Science, Marsfield NSW 2122, Australia}
\altaffiltext{7}{Bolton Fellow} \altaffiltext{8}{
INAF-Osservatorio Astronomico di Roma, via Frascati 33, I-00040
Monteporzio Catone (RM), Italy} \altaffiltext{9}{CSIRO Astronomy
and Space Science, Locked Bag 194, NSW 2390, Narrabri, Australia}

\begin{abstract}
GRB~130925A is a peculiar event characterized by an extremely long
 gamma-ray duration ($\approx$7~ks), as well as dramatic flaring in the
X-rays for $\approx$20~ks. After this period, its X-ray afterglow
shows an atypical soft spectrum with photon index $\Gamma$$\sim$4,
as observed by {\it Swift} and {\it Chandra}, until $\approx 10^7$
s, when {\it XMM-Newton} observations uncover a harder spectral
shape with $\Gamma$$\sim$2.5, commonly observed in GRB afterglows.
We find that two distinct emission components are needed to
explain the X-ray observations: a thermal component, which
dominates the X-ray emission for several weeks, and a non-thermal
component, consistent with  a typical afterglow. A forward shock model well describes the broadband
(from radio to X-rays) afterglow spectrum at various epochs. It
requires an ambient medium with a very low density wind profile,
consistent with that expected from a low-metallicity blue
supergiant (BSG). The thermal component has a remarkably constant
size and a total energy consistent with those expected by a hot
cocoon surrounding the relativistic jet.  We argue that the
features observed in this GRB (its ultralong duration, the thermal
cocoon, and the low density wind environment) are associated with
a low metallicity BSG progenitor and, thus, should characterize
the class of ultralong GRBs.
\end{abstract}



\maketitle

\section{Introduction}

Gamma-ray bursts (GRBs) are traditionally divided into two classes
based on the properties of the observed gamma-ray emission: short
duration ($<$2~s) hard spectrum and long duration ($>$2~s) soft
spectrum bursts \citep{ck93}. Between 2010 and 2012, the {\it
Swift} mission \citep{swift04} discovered three unusually
long-lasting stellar explosions that could represent a previously
unrecognized class of high-energy transients. These events, dubbed
ultra-long GRBs, persist for hours, a period up to 100 times
longer than typical GRBs. \citet{gendre13} proposed that their
unusual long duration may reflect the physical size of their
stellar progenitor, likely a low-metallicity blue supergiant
(BSG). This scenario is favored because the associated low mass loss rate
retains the outer stellar layers, that
can then continuously supply
mass to the central engine over a duration of $>$10,000~s. In fact,
accretion timescales can be crudely estimated as the
free-fall time of the external layers \citep{kumar08,qk12},
$t_{ff} \approx 10^4 R_{12}^{3/2} M_{50}^{-1/2}$ s, where the mass
$M= 50~M_{50}$\,$M_{\odot}$, and radius $R=10^{12}R_{12}$\,cm are
typical of a BSG with low  metallicity \citep{heger03,wh12}. The
lack of any bright supernova component \citep{levan14} further
supports the idea of a progenitor system different from the
compact Wolf-Rayet (WR) star, which gives birth to standard long GRBs
\citep{wb06}.

In the scenario proposed by \citet{gendre13}, ultra-long GRBs
could represent  the closest link ever discovered between GRBs and
the rare Population III stars \citep{abel02,bromm02}, which end
their lives as BSGs with massive hydrogen envelopes \citep{w02}.
Ultra-long GRBs in the local Universe offer us the unique
opportunity to study the explosion mechanisms of the most distant
stellar explosions \citep{mr10,suwa11}: those happening during the
`Cosmic Dawn' \citep{bl01}.

In this Letter, we present the results of our X-ray and radio
monitoring campaign of the recently discovered ultra-long
GRB~130925A. Besides its extreme long  gamma-ray duration
($\approx$7 ks), this event was also characterized by an unusually
soft X-ray spectrum during its afterglow phase, accompanied by a
hard-to-soft-to-hard spectral evolution in time. Contrary to
previous works, we argue that the atypical afterglow behavior of
this ultra-long GRB is closely related to its atypical stellar
progenitor.

\begin{figure}[!t]
\includegraphics[scale=0.45]{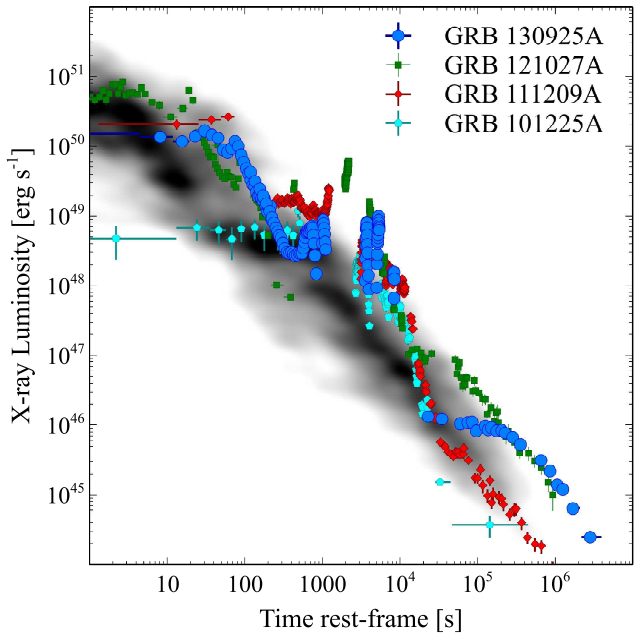}
\caption{X-ray luminosity light curves for ultra-long GRBs:
GRB101225A at $z$=0.84, GRB111209A at $z$=0.67, GRB121027A at
$z$=1.773 and GRB~130925A at $z$=0.35. The gray area shows the
light curves of $\sim$200 {\it Swift} long GRBs with measured
redshift. The canonical (steep-flat-normal) afterglow decay can be
recognized. Ultra-long GRBs display continuous prompt emission
activity up to late times, when standard long GRBs already follow
the ``normal'' afterglow decay phase. } \label{fig:lc}
\end{figure}

\section{Data analysis}

\subsection{Observations and data reduction}

GRB~130925A triggered the {\it Swift} Burst Alert Telescope (BAT)
on 2013 September 25 at 04:11:24 UT, which we refer to as $T_0$.
Its prompt gamma-ray phase was also detected at earlier times by
{\it INTEGRAL} at $T_0$-718~s \citep{savchenko13}, by the {\it
Fermi} Gamma-Ray Burst Monitor at $T_0$-15~m
\citep{fitzpatrick13}, and by Konus-Wind \citep{golenetskii13}.
The GRB emission showed several strong bursts visible by BAT up to
$\approx$7~ks, as observed in other ultra-long GRBs. A comparison
between GRB~130925A, ultra-long GRBs, and long GRBs is shown in
Fig.\ref{fig:lc}.


Pointed observations with the {\it Swift} X-ray Telescope (XRT)
began at $T_0$+151~s, revealing a bright and highly variable X-ray
afterglow. Strong flaring episodes were detected until
T$_0$+6~hrs, after which the afterglow exhibited a smooth
power-law decay (Fig.\ref{fig:lc}). Follow-up observations with
the XRT lasted for $\sim$6 months, for a total net exposure of
425~ks in Photon Counting mode. XRT data were processed using the
XRT Data Analysis Software (XRTDAS; ver.~12.9.3) distributed
within HEASOFT. We used the latest release of the XRT Calibration
Database and followed a standard reduction procedure
\cite[see][]{evans+09}. We also analyze a {\it Chandra}
observation \citep[PI:~E.~Bellm, see][]{bellm+13}, performed at
$T_0$+11 days, for a total net exposure of 44 ks. {\it Chandra}
data were reduced with the CIAO v.4.6 and the relevant calibration
files.

In order to characterize the late-time afterglow evolution, a
dedicated Target of Opportunity observation with {\it XMM-Newton}
(PI:~L.~Piro) was carried out on December~26 ($T_0$ + 3 months)
for 100~ks. {\it XMM} data were reduced using SAS version 13.5.0.
After applying standard filtering criteria and removing time
intervals with high flaring background activity, the total net
exposure is 85~ks.

The optical/IR counterpart of GRB~130925A was detected by GROND
\citep{Sudilovsky+13}, and later localized by the HST to lie 0.12\arcsec~
from the galaxy nucleus \citep{tanvir+13}. The red and relatively faint afterglow
suggests that GRB~130925A was a highly extinguished event.
Spectroscopic observations of the
underlying host galaxy measured a redshift $z$=0.347
\citep{vreeswijk+13}.

Radio observations with the Australia Telescope Compact Array
(ATCA) detected a source at a position consistent with the X-ray
and optical localizations. These observations were carried out
into three campaigns: one in October 2013 \citep{bannister+13},
one in January 2014, and one in February 2014 (PI:~L.~Piro). Radio
data were calibrated and imaged using standard procedures within
the MIRIAD data reduction package \citep{sault+95}.

\begin{table*}
\caption{Results of the spectral fits\tablenotemark{a}\label{tab:fit}}
\begin{tabularx}{\textwidth}{l X X c X X X c X X c c c X}
\hline \hline
      &
\multicolumn{2}{c}{Power Law} & &
\multicolumn{3}{c}{Power Law (N$_H$ linked)} & &
\multicolumn{6}{c}{Power Law (N$_H$ and $\Gamma$ linked)+ Black Body}\\
\cline{2-3}
\cline{5-7}
\cline{9-14}
 Time~interval &
 $N_H$ & $\Gamma$ &  &
 $N_H$ & $\Gamma$ & $\chi^2$/dof &  &
 $N_H$ & $\Gamma$ &   kT$_{BB}$  &
 $L_{44}^{BB}$ &  $R_{11}^{BB}$  & $\chi^2$/dof \\
\hline
E1: 150-500~s &
$1.36^{+0.05}_{-0.05}$ & $1.72^{+0.03}_{-0.03}$ & &
$1.38^{+0.05}_{-0.05}$ & $1.73^{+0.03}_{-0.03}$ & 852/820 & &
$1.80^{+0.15}_{-0.15}$ & $1.73^{+0.05}_{-0.08}$ & $1.4^{+0.3}_{-0.3}$ &
$2.4^{+0.9}_{-0.9}$$\times$10$^3$ & 1.2$\pm0.6$  & 806/817 \\
E2: 500-700~s &
$1.50^{+0.10}_{-0.10}$ & $1.71^{+0.06}_{-0.06}$ & &
-- & $1.67^{+0.04}_{-0.04}$ & -- & &
-- & -- & $1.30^{+0.15}_{-0.15}$ &
$2.6^{+0.6}_{-0.6}$$\times$10$^3$ & 1.4$\pm0.3$  & -- \\
E3: 1150-1340~s &
$1.40^{+0.10}_{-0.10}$ & $1.9^{+0.10}_{-0.10}$ & &
$1.40^{+0.10}_{-0.10}$ & $1.9^{+0.10}_{-0.10}$ & 71/83 & &
$1.4^{+0.2}_{-0.2}$ & $2.1^{+0.3}_{-0.3}$ &  $1.5^{+0.6}_{-0.6}$ &
 $7^{+7}_{-7}$$\times$10$^2$ & $0.6\pm0.6$  & 65/81 \\

A1: 20-300~ks & $2.30^{+0.10}_{-0.10}$ & $3.5^{+0.06}_{-0.06}$ & &
$2.10^{+0.10}_{-0.10}$ & $3.4^{+0.10}_{-0.10}$ & 552/429 & &
$1.40^{+0.10}_{-0.10}$ & $2.4^{+0.2}_{-0.2}$ &
$0.5^{+0.03}_{-0.03}$ &
23$\pm$3 & 1.00$\pm$0.10 & 471/422 \\
A2: 300-700~ks &
$2.7^{+0.2}_{-0.2}$ & $4.6^{+0.2}_{-0.2}$ & &
-- & $4.10^{+0.10}_{-0.10}$ & -- & &
-- & -- &  $0.45^{+0.03}_{-0.03}$ &
10.0$\pm$1.0 & 0.90$\pm$0.10  & -- \\
A3: 0.7-2~Ms &
$1.9^{+0.2}_{-0.2}$ & $4.0^{+0.2}_{-0.2}$ & &
-- & $4.2^{+0.2}_{-0.2}$ & -- & &
-- & -- &  $0.34^{+0.04}_{-0.04}$ &
4.2$\pm$0.8 & 0.9$\pm$0.2 & -- \\
A4\tablenotemark{b}: 0.95-1~Ms &
$1.90^{+0.10}_{-0.10}$ & $3.80^{+0.10}_{-0.10}$ & &
-- & $4.0^{+0.10}_{-0.10}$ & -- & &
-- & -- &  $0.35^{+0.03}_{-0.03}$ &
3.4$\pm$0.4 & 0.80$\pm$0.10  & -- \\
A5: 3-8~Ms &
$1.3^{+0.4}_{-0.4}$ & $3.0^{+0.4}_{-0.4}$ & &
-- & $3.8^{+0.5}_{-0.5}$ & -- & &
-- & -- &  $0.30^{+0.10}_{-0.10}$ &
0.3$\pm$0.2 & $0.4^{+0.6}_{-0.3}$  & -- \\
A6\tablenotemark{c}: 0.78-8~Ms &
$1.10^{+0.14}_{-0.14}$ & $2.50^{+0.10}_{-0.10}$ & &
-- & $3.5^{+0.2}_{-0.2}$ & -- & &
-- & -- &  $0.23^{+0.05}_{-0.05}$ &
0.17$\pm$0.10 & 0.4$\pm$0.3  & -- \\
\hline
\vspace{-10pt}
\tablenotetext{1}{NOTES:
For the power-law model, we report the absorbing column $N_H$ in units of $10^{22}$\,cm$^{-2}$,
and the photon index $\Gamma$.
For the blackbody model, we report the black-body temperature $kT_{BB}$ in units of keV,
luminosity $L_{BB}$ in units of $10^{44}$\,erg\,s$^{-1}$,
and radius $R_{BB}$ in units of $10^{11}$\,cm.}
\tablenotetext{2}{Chandra}
\tablenotetext{3}{XMM-Newton}
\end{tabularx}
\end{table*}

\subsection{X-rays}

A detailed analysis of the early X-ray flares is presented in
\cite{evans14}; here we focus on the late ($>$20~ks) afterglow
emission. The X-ray emission in the 0.3-10 keV energy band decays
as a simple power-law function, $f_X \propto t^{-\alpha}$, with
slope $\alpha$=0.82, steepening to $\alpha$=1.32 after $\sim$300
ks. The hardness ratio light curve displays significant variations
up to late times, which is unusual for standard GRB afterglows. In
order to quantify the spectral evolution, we performed a
time-resolved spectral analysis using XSPEC v.12.8.1
\citep{xspec}. Our results are summarized in Table~1 (spectra
A1-A6). The X-ray spectra were described by an absorbed power-law
model. The Galactic absorption component was kept fixed at the
value of $1.66\times 10^{20}$ cm$^{-2}$. A redshifted absorption
component, modeling the host intrinsic absorption, was initially
left free to vary. The afterglow spectrum shows a hard-to-soft
(from $\Gamma$$\sim$3.5 to $\sim$4.6) followed by a soft-to-hard
(from $\Gamma$$\sim$4.6 to $\sim$2.5) evolution. The spectral
fits, although acceptable, yield unphysical variations of the
absorbing column, correlated with the evolution of the power-law
index (Table~1, col.~2-3). We, therefore, linked the intrinsic
absorption between the different spectra (Table~1, col.~4-6). The
resulting fit is poor ($\chi^2$=552 for 429 dof), mainly because
the soft power-law spectrum underestimates the flux above
$\sim$3~keV.

A steep non-thermal spectrum (3.3$<\Gamma<$4.4) is atypical in GRB
afterglows, and is generally seen as an indication of a thermal
component which, over the limited 0.3-10 keV energy bandpass,
cannot be fully resolved. Each spectrum was then fit by adding a
black body component to the simple non-thermal power-law. The
improvement of \chisq~is highly significant. However, despite the
good statistics of the {\it Swift} and {\it Chandra} spectra, the
spectral parameters were not well constrained,  with an error
range that included the blackbody temperatures and spectral
indices found by Bellm et al. (2014). This is mainly because the
flux of the blackbody component is comparable to, or even larger
than, the power-law flux and dominates the emission below 3~keV.

In this respect, the late {\it XMM} observation was crucial to
constrain the model. The high throughput of the telescope  allowed
us to gather enough counts at late times, when the thermal
component had shifted to lower temperatures, and the two emission
components could be better disentagled. The {\it XMM} spectrum can
be described by a power law with $\Gamma=2.50\pm0.10$, rather
common in GRB afterglows. The blackbody component is still
significantly detected ($\Delta$\chisq=15.3 corresponding to a
chance probability of $6.8\times10^{-4}$) with temperature
$kT=0.25\pm0.04$~keV.

Based on these results, we tested whether the observed X-ray emission
could be described by an underlying non-thermal afterglow,
and a dominating, highly variable thermal component.
The spectra were simultaneously fit by linking the absorbing column
and the power-law photon index, and by letting the black-body parameters
free to vary (Table~1, col. 7-12). Compared to the reference model
(col. 4-6), the addition of the blackbody improves the fit
at a very high level, yielding a $\Delta$\chisq=81 for 7 additional
parameters corresponding to a chance probability of $<<10^{-8}$.

Motivated by the results derived at $t\gtrsim20$\,ks, we also
searched for a thermal component at earlier times,
excluding the periods dominated by the X-ray flares.
The results are reported in Table~1 (spectra E1-E3).
The early-time spectra are dominated by the non-thermal emission,
but consistent with the presence of a black body component
with temperature $kT_{BB}\approx 1$\,keV.

\subsection{Radio}

\begin{table}[b]
\begin{center}
\caption{Radio observations of GRB 130925A}
\begin{tabular}{lcccc}
\hline
\hline
Date    &   5.5\,GHz      &  9\,GHz        &   17\,GHz       &    19\,GHz       \\
        &   [$\mu$Jy]      &  [$\mu$Jy]      &   [$\mu$Jy]      &    [$\mu$Jy]      \\
\hline
Oct 09, 2013 &   ---         & 161$\pm$ 24  &   146$\pm$ 22 &   109$\pm$28   \\
Oct 14, 2013 &   ---         & 162$\pm$19   &   ---         &   ---          \\
Oct 15, 2013 &   ---         & ---          &   170$\pm$31  &   169$\pm$43   \\
Jan 04, 2014 &   165$\pm$ 31 & 245$\pm$30   &    58$\pm$18  &   175$\pm$19 \\
\hline
\end{tabular}
\label{tab:fluxes}
\end{center}
\end{table}

The full set of ATCA measurements is listed in
Table~\ref{tab:fluxes}. Flux densities and corresponding 1$\sigma$
errors were obtained from the task {\it maxfit} in MIRIAD. In the
first two campaigns (October 2013, and January 2014) the source
was localized with typical uncertainty of 0.9-1.3\arcsec.  To
further improve the positional accuracy of the target, another 17
and 19-GHz follow-up was carried out in February 2014 with the
ATCA in its most extended array configuration (6D: maximum
baseline length 6\,km). Notwithstanding  poor observing conditions
we were able to obtain a weak (S/N=3.3) detection of the target at
17 GHz. The best positional accuracy was obtained by fitting a 2-D
Gaussian model to the target image at 17~GHz with the MIRIAD task
{\it imfit}. The resulting position is: RA=02:44:42.949,
Dec=-26:09:11.090 with (1$\sigma$) errors $\Delta$RA=0.106\arcsec,
and $\Delta$Dec=0.621\arcsec.

The radio source remains nearly constant in brightness (within the
uncertainties) over a period of $\sim$4 months. A comparison with
the HST images, which we downloaded from the public archive, shows
that the radio position is consistent with the faint transient
reported by \cite{tanvir+13} and offset from the galaxy nucleus.
The probability of a chance alignment for a source this brightness
is negligible ($P$$\approx$3\ee{-5}), and we conclude that the
radio source is the GRB afterglow. Its relatively constant flux
suggests that the blastwave is expanding into a circumburst medium
with a wind-like density profile, $\rho(r)\propto r^{-2}$.

\section{Discussion}


Two salient features characterize GRB~130925A: an extreme long
duration, and a very steep X-ray spectrum. As shown in Figure~1,
an intense and persistent flaring activity dominates the emission
for the first six hours while, over this period, typical long GRBs
already entered the normal afterglow phase. After the flaring
ceases, the X-ray emission displays an unusually steep spectrum.
The scenario that we have tested envisions the presence of two
components contributing to the X-ray emission: a thermal
component, well described by a black body emission with constant
radius, decreasing temperature and luminosity; and an underlying
non-thermal component. In the following we discuss their
properties, and their possible origin.


\subsection{External shock into a low-density wind environment}

\begin{figure}[!t]
\centering
\includegraphics[angle=0,scale=0.4]{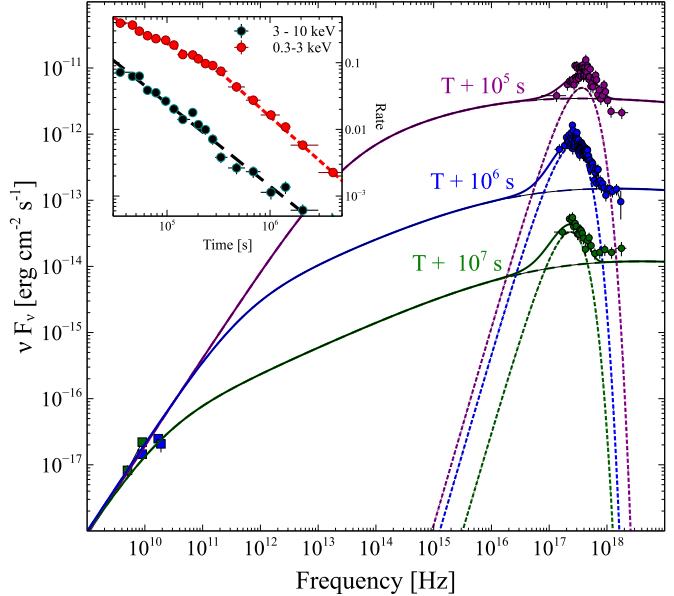}
 \caption{Broadband afterglow spectra at three different epochs.
The best fit model consists of a standard forward shock in a wind
environment (solid line), and a black body component (dashed
line). The best fit parameters are $E_{k,\rm iso}=10^{53}$ erg,
$A_{*}=3.6\times 10^{-3}$, $\epsilon_e$=0.16, $\epsilon_B=0.33$,
p=2.25. The light curves in the 0.3-3 keV and 3-10 keV bands are
presented in the inset.} \label{fig:spectra}
\end{figure}

The non-thermal emission is well-described by a power-law with spectral slope
$\beta=\Gamma-1=1.4\pm0.2$, consistent with the spectral
indices observed in GRB afterglows
\citep{depasquale06,willingale07}.  The X-ray light curve
above 3 keV (inset of Fig.~\ref{fig:spectra}), where the non-thermal component
dominates, shows a power-law temporal decay with $\alpha=1.20\pm 0.05$,
consistent with the closure relation
$\alpha=(3\beta-1)/2$ for $\nu_X>\nu_c$.
This evidence motivated us in modelling the
X-ray non-thermal component and the radio data with a standard
forward shock model \citep{gs02}. The broad band spectra from
radio to X-rays at three epochs were fitted simultaneously,
yielding a good agreement with a forward shock expanding in a wind-like
environment. The best fit model and
parameters are presented in Fig.~\ref{fig:spectra}.
The blastwave energy $E_k\approx10^{53}$\,erg is comparable to the observed
gamma-ray energy $E_{\gamma,iso} \approx 1.5 \times 10^{53}$\,erg,
which implies a high radiative efficiency of the prompt emission mechanism.
The tenuous, wind driven medium derived from the fit
implies a low mass loss rate of $\Mdot\approx 3.6 \times 10^{-8}$~\msun\,yr$^{-1}$
for a wind velocity of $v_w=10^3 {\rm\ km\ s^{-1}}$, consistent with
a very low-metallicity BSG progenitor \citep{vkl01,kudritzky02}.
From the lack of jet-break signature in the 3-10~keV light curve,
we derive a lower limit on the jet-break time $t_j\gtrsim 90$\,d,
and a jet opening \mbox{angle $\theta_j$\,$\gtrsim$\,2\,deg.}

\subsection{A hot cocoon}

\begin{figure}[!t]
\centering
\includegraphics[angle=0,scale=0.5]{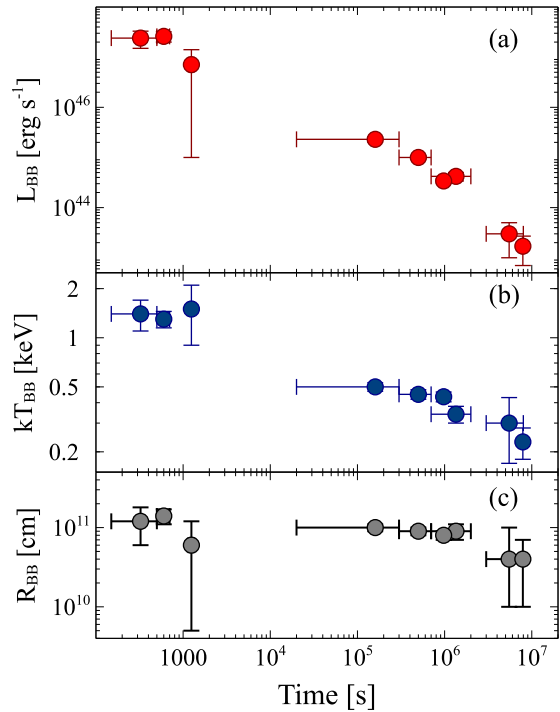}
 \caption{Parameters of the Black body component. Luminosity (a), temperature (b) and radius of the emission region (c).}
\label{fig:bbody}
\end{figure}

The luminosity, temperature, and apparent radius of the blackbody
component are plotted in Fig.~\ref{fig:bbody} as  functions of
time. The luminosity declines as $L_{BB} \propto t^{-0.9}$, while
the temperature exhibits a slower decreasing trend, $T \propto
t^{-0.2}$. The apparent radius is $R_{BB}\approx 10^{11}$ cm,
showing that, within the uncertainties, the size of the emitting
source remained remarkably constant in spite of the large
variation in luminosity.

%

Three main mechanisms can produce a thermal component in GRBs,
namely a shock break out, the jet photosphere,
and a hot cocoon.
Shock breakouts are characterized by durations $<<10,000$s and peak X-ray
luminosities from $\approx 10^{44}$ erg s$^{-1}$ to $10^{46}$ erg
s$^{-1}$ \citep{eb92, campana+06}, not consistent with the
long timescale and large luminosity observed in GRB~130925A.



Bright thermal emission from the fireball photosphere may emerge
during the prompt gamma-ray phase. This high energy photospheric
component is associated to the optically thick plasma of a
relativistic jet, and decays in luminosity and temperature as a
power-law in time \citep{rp09}. It may still be detectable in the
soft X-rays a few hundreds seconds after the burst \citep{ss12,fw13}
and, in principle, can continue on much longer time scales if the
jet continues to be powered. \cite{wong+14} showed that for a BSG
this can be indeed the case: the fall-back of the external layers
onto the central black hole yields an accretion rate
$\Mdot$\,$\approx$\,10~t$^{-5/3}$\,\msun~yr$^{-1}$, and a
corresponding  jet luminosity
$L_{jet}$\,$\approx$\,2$\times$10$^{49}$\,$\eta_{-1}$\,$t_{3}^{-5/3}$\,erg\,s$^{-1}$.
Here  $\eta=0.1~\eta_{-1}$ is the mass to energy conversion
efficiency, and $t= 1000\ t_3~{\rm s}$. The photospheric radius is
$r_{ph}=5.8\ 10^{11} L_{jet,51} \Gamma_2^{-3}$~cm
\citep{anp91,peer+12}, where $\Gamma$=100\,$\Gamma_2$ is the jet
Lorentz factor. Therefore, a photospheric emission with constant
$r_{ph}\approx 10^{11}$~cm requires $\Gamma_2\approx$
0.5\,$\eta_{-1}^{1/3} t_3^{-5/9}$ from a few hundreds seconds to
10$^7$\,s.
This fine coupling $\Gamma\propto L_{jet}^{1/3}$ required to keep a constant photospheric
radius seems somewhat contrived, although it cannot be excluded \citep[e.g.][]{fan+12}

Let us now discuss the association of the blackbody with a hot
plasma cocoon. 
The cocoon develops inside the star by the
interaction of the jet with the stellar layers, and eventually
breaks out at the stellar surface when the jet emerges
\citep{lazzati05}.  \cite{starling+12} have proposed that the black body component  found in a few GRBs during the steep decay phase of the X-ray light curve can be associated, at least in one event, to   a relativistically expanding hot plasma cocoon. This component is short-lived ($\ls 1,000$ s) and its radius is rapidly increasing with time.
 On the contrary, our observations exhibit a long-lasting black body emission with a constant radius, indicating that the cocoon does
not expand. Thus   some process must confine it as it emerges at
the stellar surface. A promising mechanism is magnetic confinement
\citep{k99}. For instance, the toroidal component of the magnetic
field could be advected into the inner part of the cocoon
\citep{lb13}, suppressing the plasma expansion across the magnetic
field lines and confining it around the jet. In this scenario, the
transverse size of the cocoon when it emerges at the stellar
surface should be similar to the jet opening angle, $\theta_c
\approx \theta_j \approx R_{BB}/R_*$, where $R_*$ is the radius of
the progenitor star. The limit on the jet opening angle derived
previously implies $R_*\ls 3\times 10^{12}$cm, consistent with
 the typical radii of a BSG producing an accretion disk after
the collapse \citep{wh12,kashiyama+13}. The energy of the baryons
entrained in the cocoon can be derived as $E_{th,b}\approx 3 M_c/
2 m_p kT \approx 3\times 10^{47}$\,erg. Here $M_c$ is the stellar
mass contained within the volume escavated by the jet, that is
$V_c \sim \pi R_{BB}^2 R_*\approx 10^{35}$ \cm{3}. This energy is
several orders of magnitudes lower than the energy of the
accretion-powered jet, $E_{jet}\approx 5 \times
10^{53}$\,$\eta_{-1}$ \,erg \citep{wong+14}, thus, if the jet is
Poynting-flux dominated, its magnetic field can easily confine the
plasma of the cocoon.

The blackbody energetics, $E_{BB}$=1.5$\times 10^{51}$~erg, are safely below the energy
produced by a jet piercing through a BSG, $E_{c} \approx 10^{52}$\,erg \citep{kashiyama+13},
and are much larger  than the energy of the baryons computed above.
The cocoon is therefore radiation dominated, and its temperature  at the instant of the break out
can be estimated as $kT_{BB} \lesssim ( E_{BB}/V_c)^{1/4}\approx$~1~keV, consistent with that observed.
The cocoon's cooling time, $t_c \approx E_{BB}/L_{BB} \lesssim 10^{4}$\,s, is however much shorter
than the duration of the thermal emission.
Thus, the cocoon must be continuously energized by a
fraction $\eta_{th}$ of the jet energy which, as mentioned above, follows
a $L_{jet}\propto t^{-5/3}$.  The black body luminosity can be expressed
as $L_{BB}\approx \eta_{th}  L_{jet}$. In our case $\eta_{th}\approx 0.01$.
The flatter decay slope of the thermal component can be
accounted by a slow increase of $\eta_{th}$ when the jet energy
decreases.


\section{Conclusions}

We conclude that the features observed in this GRB (its
extremely long duration, the thermal X-ray spectrum, and the low density wind
environment) are associated with a low metallicity BSG progenitor, and
could characterize the class of ultralong GRBs as a whole.
The fallback of the stellar outer layers supports a long duration jet,
which entrains a large mass of baryons in a hot cocoon. This mass is larger than that
available for a WR star, the proposed progenitor of long GRBs.
Due to the larger baryon loading, it is likely that the
jet Lorentz factor at the surface of the BSG is
substantially lower than the equivalent for a WR
progenitor, eventually favoring a larger efficiency $\eta_{th}$.
Both these effects boost the emission from the thermal cocoon in
ultra-long GRBs as opposed to the case of standard long GRBs.
With a typical rest-frame temperature of $kT_{BB} \sim 0.5$ keV
the thermal component could not be detected in the other
three presently known ultralong GRBs, all lying at larger $z$.

PopIII star are expected to end their life as BSG \citep{w02,naka12}, thus their  explosions should be characterized by features similar to those observed in this ultralong GRB.
Pop~III stars are first formed in the early universe where the low
value of the metallicity ($Z$$<$$Z_{cr}\approx 10^{-4}$) favors
the formation of large ($>100$\msun) collapsing gas clouds.
Chemical enrichment following the first stars explosions proceed
inhomogeneously, thus Population~III stars can continue to form
until late epochs, provided that gas pockets of sufficiently low
metallicity can be preserved during cosmic evolution. Could  the
progenitor of GRB130925A be a rare Pop~III star? Various authors
\cite[e.g.][]{tfs07} have shown that a sizeable fraction of
Pop~III stars can form down to $z\approx2.5$ and, possibly, this
formation could extend into the local Universe. The mass loss rate
derived from the afterglow fit is suggestive of a metal-poor star.
However, it cannot lead to quantitative measurements, as the mass
loss is not uniquely dependent upon metallicity. The low redshift
and the high dust content of the host galaxy represent an
environment more typical of a Population II star, although only
deep spectroscopic observations of the GRB birthsite may
ultimately elucidate the nature of the progenitor.

\section{Acknowledgements}
We acknowledge useful discussion with A. Chieffi, C. Macculi and R. Salvaterra.
We thank the XMM team for the support in carrying out the TOO
observation.

\bibliographystyle{aa}

\begin{thebibliography}{50}
\expandafter\ifx\csname natexlab\endcsname\relax\def\natexlab#1{#1}\fi


\bibitem[{{Abel} {et~al.}(2002){Abel}, {Bryan}, \& {Norman}}]{abel02}
{Abel}, T., {Bryan}, G.~L., \& {Norman}, M.~L. 2002, Science, 295, 93

\bibitem[{{Abramowicz} {et~al.}(1991){Abramowicz}, {Novikov}, \&
  {Paczynski}}]{anp91}
{Abramowicz}, M.~A., {Novikov}, I.~D., \& {Paczynski}, B. 1991, \apj, 369, 175

\bibitem[{{Arnaud}(1996)}]{xspec}
{Arnaud}, K.~A. 1996, in Astronomical Society of the Pacific Conference Series,
  Vol. 101, Astronomical Data Analysis Software and Systems V, ed. G.~H.
  {Jacoby} \& J.~{Barnes}, 17

\bibitem[{{Bannister} {et~al.}(2013){Bannister}, {Hancock}, {Kulkarni},
  {Horesh}, {Zauderer}, {Murphy}, \& {Gaensler}}]{bannister+13}
{Bannister}, K., {Hancock}, P., {Kulkarni}, S., {Horesh}, A., {Zauderer}, A.,
  {Murphy}, T., \& {Gaensler}, B. 2013, The Astronomer's Telegram, 5531, 1

\bibitem[{{Barkana} \& {Loeb}(2001)}]{bl01}
{Barkana}, R. \& {Loeb}, A. 2001, \physrep, 349, 125

\bibitem[{{Bellm} {et~al.}(2014){Bellm}, {Barri{\`e}re}, {Bhalerao}, {Boggs},
  {Cenko}, {Christensen}, {Craig}, {Forster}, {Fryer}, {Hailey}, {Harrison},
  {Horesh}, {Kouveliotou}, {Madsen}, {Miller}, {Ofek}, {Perley}, {Rana},
  {Reynolds}, {Stern}, {Tomsick}, \& {Zhang}}]{bellm+13}
{Bellm}, E.~C., {et~al.} 2014, \apjl, 784, L19

\bibitem[{{Bromm} {et~al.}(2002){Bromm}, {Coppi}, \& {Larson}}]{bromm02}
{Bromm}, V., {Coppi}, P.~S., \& {Larson}, R.~B. 2002, \apj, 564, 23

\bibitem[{{Campana} {et~al.}(2006){Campana}, {Mangano}, {Blustin}, {Brown},
  {Burrows}, {Chincarini}, {Cummings}, {Cusumano}, {Della Valle}, {Malesani},
  {M{\'e}sz{\'a}ros}, {Nousek}, {Page}, {Sakamoto}, {Waxman}, {Zhang}, {Dai},
  {Gehrels}, {Immler}, {Marshall}, {Mason}, {Moretti}, {O'Brien}, {Osborne},
  {Page}, {Romano}, {Roming}, {Tagliaferri}, {Cominsky}, {Giommi}, {Godet},
  {Kennea}, {Krimm}, {Angelini}, {Barthelmy}, {Boyd}, {Palmer}, {Wells}, \&
  {White}}]{campana+06}
{Campana}, S., {et~al.} 2006, \nat, 442, 1008

\bibitem[{{de Pasquale} {et~al.}(2006){de Pasquale}, {Piro}, {Gendre}, {Amati},
  {Antonelli}, {Costa}, {Feroci}, {Frontera}, {Nicastro}, {Soffitta}, \& {in't
  Zand}}]{depasquale06}
{de Pasquale}, M., {et~al.} 2006, \aap, 455, 813

\bibitem[{{Ensman} \& {Burrows}(1992)}]{eb92}
{Ensman}, L. \& {Burrows}, A. 1992, \apj, 393, 742

\bibitem[{{Evans} {et~al.}(2009){Evans}, {Beardmore}, {Page}, {Osborne},
  {O'Brien}, {Willingale}, {Starling}, {Burrows}, {Godet}, {Vetere}, {Racusin},
  {Goad}, {Wiersema}, {Angelini}, {Capalbi}, {Chincarini}, {Gehrels}, {Kennea},
  {Margutti}, {Morris}, {Mountford}, {Pagani}, {Perri}, {Romano}, \&
  {Tanvir}}]{evans+09}
{Evans}, P.~A., {et~al.} 2009, \mnras, 397, 1177

\bibitem[{{Evans} {et~al.}(2014){Evans}, {Willingale}, {Osborne}, {O'Brien},
  {Tanvir}, {Frederiks}, {Pal'shin}, {Svinkin}, {Lien}, {Cummings}, {Xiong},
  {Zhang}, {G{\"o}tz}, {Savchenko}, {Negoro}, {Nakahira}, {Suzuki}, {Wiersema},
  {Starling}, {Castro-Tirado}, {Beardmore}, {S{\'a}nchez-Ram{\'{\i}}rez},
  {Gorosabel}, {Jeong}, {Kennea}, {Burrows}, \& {Gehrels}}]{evans14}
{Evans}, P.~A., {et~al.} 2014, ArXiv e-prints

\bibitem[{{Fan} {et~al.}(2012){Fan}, {Wei}, {Zhang}, \& {Zhang}}]{fan+12}
{Fan}, Y.-Z., {Wei}, D.-M., {Zhang}, F.-W., \& {Zhang}, B.-B. 2012, \apjl, 755,
  L6

\bibitem[{{Fitzpatrick}(2013)}]{fitzpatrick13}
{Fitzpatrick}, G. 2013, GRB Coordinates Network, 15255, 1

\bibitem[{{Friis} \& {Watson}(2013)}]{fw13}
{Friis}, M. \& {Watson}, D. 2013, \apj, 771, 15

\bibitem[{{Gehrels} {et~al.}(2004){Gehrels}, {Chincarini}, {Giommi}, {Mason},
  {Nousek}, {Wells}, {White}, {Barthelmy}, {Burrows}, {Cominsky}, {Hurley},
  {Marshall}, {M{\'e}sz{\'a}ros}, {Roming}, {Angelini}, {Barbier}, {Belloni},
  {Campana}, {Caraveo}, {Chester}, {Citterio}, {Cline}, {Cropper}, {Cummings},
  {Dean}, {Feigelson}, {Fenimore}, {Frail}, {Fruchter}, {Garmire}, {Gendreau},
  {Ghisellini}, {Greiner}, {Hill}, {Hunsberger}, {Krimm}, {Kulkarni}, {Kumar},
  {Lebrun}, {Lloyd-Ronning}, {Markwardt}, {Mattson}, {Mushotzky}, {Norris},
  {Osborne}, {Paczynski}, {Palmer}, {Park}, {Parsons}, {Paul}, {Rees},
  {Reynolds}, {Rhoads}, {Sasseen}, {Schaefer}, {Short}, {Smale}, {Smith},
  {Stella}, {Tagliaferri}, {Takahashi}, {Tashiro}, {Townsley}, {Tueller},
  {Turner}, {Vietri}, {Voges}, {Ward}, {Willingale}, {Zerbi}, \&
  {Zhang}}]{swift04}
{Gehrels}, N., {et~al.} 2004, \apj, 611, 1005

\bibitem[{{Gendre} {et~al.}(2013){Gendre}, {Stratta}, {Atteia}, {Basa},
  {Bo{\"e}r}, {Coward}, {Cutini}, {D'Elia}, {Howell}, {Klotz}, \&
  {Piro}}]{gendre13}
{Gendre}, B., {et~al.} 2013, \apj, 766, 30

\bibitem[{{Golenetskii} {et~al.}(2013){Golenetskii}, {Aptekar}, {Frederiks},
  {Pal'Shin}, {Oleynik}, {Ulanov}, {Svinkin}, \& {Cline}}]{golenetskii13}
{Golenetskii}, S., {Aptekar}, R., {Frederiks}, D., {Pal'Shin}, V., {Oleynik},
  P., {Ulanov}, M., {Svinkin}, D., \& {Cline}, T. 2013, GRB Coordinates
  Network, 15260, 1

\bibitem[{{Granot} \& {Sari}(2002)}]{gs02}
{Granot}, J. \& {Sari}, R. 2002, \apj, 568, 820

\bibitem[{{Heger} {et~al.}(2003){Heger}, {Fryer}, {Woosley}, {Langer}, \&
  {Hartmann}}]{heger03}
{Heger}, A., {Fryer}, C.~L., {Woosley}, S.~E., {Langer}, N., \& {Hartmann},
  D.~H. 2003, \apj, 591, 288

\bibitem[{{Kashiyama} {et~al.}(2013){Kashiyama}, {Nakauchi}, {Suwa}, {Yajima},
  \& {Nakamura}}]{kashiyama+13}
{Kashiyama}, K., {Nakauchi}, D., {Suwa}, Y., {Yajima}, H., \& {Nakamura}, T.
  2013, \apj, 770, 8

\bibitem[{{Komissarov}(1999)}]{k99}
{Komissarov}, S.~S. 1999, \mnras, 308, 1069

\bibitem[{{Kouveliotou} {et~al.}(1993){Kouveliotou}, {Meegan}, {Fishman},
  {Bhat}, {Briggs}, {Koshut}, {Paciesas}, \& {Pendleton}}]{ck93}
{Kouveliotou}, C., {Meegan}, C.~A., {Fishman}, G.~J., {Bhat}, N.~P., {Briggs},
  M.~S., {Koshut}, T.~M., {Paciesas}, W.~S., \& {Pendleton}, G.~N. 1993, \apjl,
  413, L101

\bibitem[{{Kudritzki}(2002)}]{kudritzky02}
{Kudritzki}, R.~P. 2002, \apj, 577, 389

\bibitem[{{Kumar} {et~al.}(2008){Kumar}, {Narayan}, \& {Johnson}}]{kumar08}
{Kumar}, P., {Narayan}, R., \& {Johnson}, J.~L. 2008, \mnras, 388, 1729

\bibitem[{{Lazzati} \& {Begelman}(2005)}]{lazzati05}
{Lazzati}, D. \& {Begelman}, M.~C. 2005, \apj, 629, 903

\bibitem[{{Levan} {et~al.}(2014){Levan}, {Tanvir}, {Starling}, {Wiersema},
  {Page}, {Perley}, {Schulze}, {Wynn}, {Chornock}, {Hjorth}, {Cenko},
  {Fruchter}, {O'Brien}, {Brown}, {Tunnicliffe}, {Malesani}, {Jakobsson},
  {Watson}, {Berger}, {Bersier}, {Cobb}, {Covino}, {Cucchiara}, {de Ugarte
  Postigo}, {Fox}, {Gal-Yam}, {Goldoni}, {Gorosabel}, {Kaper}, {Kr{\"u}hler},
  {Karjalainen}, {Osborne}, {Pian}, {S{\'a}nchez-Ram{\'{\i}}rez}, {Schmidt},
  {Skillen}, {Tagliaferri}, {Th{\"o}ne}, {Vaduvescu}, {Wijers}, \&
  {Zauderer}}]{levan14}
{Levan}, A.~J., {et~al.} 2014, \apj, 781, 13

\bibitem[{{Levinson} \& {Begelman}(2013)}]{lb13}
{Levinson}, A. \& {Begelman}, M.~C. 2013, \apj, 764, 148

\bibitem[{{M{\'e}sz{\'a}ros} \& {Rees}(2010)}]{mr10}
{M{\'e}sz{\'a}ros}, P. \& {Rees}, M.~J. 2010, \apj, 715, 967

\bibitem[{{Nakauchi} {et~al.}(2012){Nakauchi}, {Suwa}, {Sakamoto}, {Kashiyama},
  \& {Nakamura}}]{naka12}
{Nakauchi}, D., {Suwa}, Y., {Sakamoto}, T., {Kashiyama}, K., \& {Nakamura}, T.
  2012, \apj, 759, 128

\bibitem[{{Pe'er} {et~al.}(2012){Pe'er}, {Zhang}, {Ryde}, {McGlynn}, {Zhang},
  {Preece}, \& {Kouveliotou}}]{peer+12}
{Pe'er}, A., {Zhang}, B.-B., {Ryde}, F., {McGlynn}, S., {Zhang}, B., {Preece},
  R.~D., \& {Kouveliotou}, C. 2012, \mnras, 420, 468

\bibitem[{{Quataert} \& {Kasen}(2012)}]{qk12}
{Quataert}, E. \& {Kasen}, D. 2012, \mnras, 419, L1

\bibitem[{{Ryde} \& {Pe'er}(2009)}]{rp09}
{Ryde}, F. \& {Pe'er}, A. 2009, \apj, 702, 1211

\bibitem[{{Sault} {et~al.}(1995){Sault}, {Teuben}, \& {Wright}}]{sault+95}
{Sault}, R.~J., {Teuben}, P.~J., \& {Wright}, M.~C.~H. 1995, in Astronomical
  Society of the Pacific Conference Series, Vol.~77, Astronomical Data Analysis
  Software and Systems IV, ed. R.~A. {Shaw}, H.~E. {Payne}, \& J.~J.~E.
  {Hayes}, 433

\bibitem[{{Savchenko} {et~al.}(2013){Savchenko}, {Beckmann}, {Ferrigno},
  {Bozzo}, {Beck}, {Borkowski}, {Gotz}, {Mereghetti}, {von Kienlin}, {Rau}, \&
  {Hurley}}]{savchenko13}
{Savchenko}, V., {et~al.} 2013, GRB Coordinates Network, 15259, 1

\bibitem[{{Sparre} \& {Starling}(2012)}]{ss12}
{Sparre}, M. \& {Starling}, R.~L.~C. 2012, \mnras, 427, 2965

\bibitem[{{Starling} {et~al.}(2012){Starling}, {Page}, {Pe'Er}, {Beardmore}, \&
  {Osborne}}]{starling+12}
{Starling}, R.~L.~C., {Page}, K.~L., {Pe'Er}, A., {Beardmore}, A.~P., \&
  {Osborne}, J.~P. 2012, \mnras, 427, 2950

\bibitem[{{Sudilovsky} {et~al.}(2013){Sudilovsky}, {Kann}, \&
  {Greiner}}]{Sudilovsky+13}
{Sudilovsky}, V., {Kann}, D.~A., \& {Greiner}, J. 2013, GRB Coordinates
  Network, 15247, 1

\bibitem[{{Suwa} \& {Ioka}(2011)}]{suwa11}
{Suwa}, Y. \& {Ioka}, K. 2011, \apj, 726, 107

\bibitem[{{Tanvir} {et~al.}(2013){Tanvir}, {Levan}, {Hounsell}, {Fruchter},
  {Cenko}, {Perley}, \& {O'Brien}}]{tanvir+13}
{Tanvir}, N.~R., {Levan}, A.~J., {Hounsell}, R., {Fruchter}, A.~S., {Cenko},
  S.~B., {Perley}, D.~A., \& {O'Brien}, P.~T. 2013, GRB Coordinates Network,
  15489, 1

\bibitem[{{Tornatore} {et~al.}(2007){Tornatore}, {Ferrara}, \&
  {Schneider}}]{tfs07}
{Tornatore}, L., {Ferrara}, A., \& {Schneider}, R. 2007, \mnras, 382, 945

\bibitem[{{Vink} {et~al.}(2001){Vink}, {de Koter}, \& {Lamers}}]{vkl01}
{Vink}, J.~S., {de Koter}, A., \& {Lamers}, H.~J.~G.~L.~M. 2001, \aap, 369, 574

\bibitem[{{Vreeswijk} {et~al.}(2013){Vreeswijk}, {Malesani}, {Fynbo}, {De Cia},
  \& {Ledoux}}]{vreeswijk+13}
{Vreeswijk}, P.~M., {Malesani}, D., {Fynbo}, J.~P.~U., {De Cia}, A., \&
  {Ledoux}, C. 2013, GRB Coordinates Network, 15249, 1

\bibitem[{{Willingale} {et~al.}(2007){Willingale}, {O'Brien}, {Osborne},
  {Godet}, {Page}, {Goad}, {Burrows}, {Zhang}, {Rol}, {Gehrels}, \&
  {Chincarini}}]{willingale07}
{Willingale}, R., {et~al.} 2007, \apj, 662, 1093

\bibitem[{{Wong} {et~al.}(2014){Wong}, {Fryer}, {Ellinger}, {Rockefeller}, \&
  {Kalogera}}]{wong+14}
{Wong}, T.-W., {Fryer}, C.~L., {Ellinger}, C.~I., {Rockefeller}, G., \&
  {Kalogera}, V. 2014, ArXiv e-prints

\bibitem[{{Woosley} \& {Bloom}(2006)}]{wb06}
{Woosley}, S.~E. \& {Bloom}, J.~S. 2006, \araa, 44, 507

\bibitem[{{Woosley} \& {Heger}(2012)}]{wh12}
{Woosley}, S.~E. \& {Heger}, A. 2012, \apj, 752, 32

\bibitem[{{Woosley} {et~al.}(2002){Woosley}, {Heger}, \& {Weaver}}]{w02}
{Woosley}, S.~E., {Heger}, A., \& {Weaver}, T.~A. 2002, Reviews of Modern
  Physics, 74, 1015

\end{thebibliography}

\end{document}